\documentclass[aps,prl,twocolumn,amsmath,amssymb,floatfix,superscriptaddress]{revtex4} 
\usepackage{graphicx}  
\usepackage{dcolumn}   
\usepackage{bm}        
\usepackage{color}
\newcommand{\Eq}[1]{Eq.~(\ref{#1})}
\newcommand{\fig}[1]{Fig.~\ref{#1}}
\newcommand{\msd}{\langle r^2(\tau)\rangle}
\newcommand{\msdt}{\msd_t}
\newcommand{\malpha}{\langle\alpha\rangle}

\begin{document}

\title{Elucidating the origin of heterogeneous anomalous diffusion
  in the cytoplasm of mammalian cells}
\author{Adal Sabri$^{1}$, Xinran Xu$^{2}$, Diego Krapf$\,^{2,3,*}$,
  and Matthias Weiss}
\email[Corresponding authors: ]{diego.krapf@colostate.edu,
  matthias.weiss@uni-bayreuth.de}
\affiliation{Experimental Physics I, University of Bayreuth, D-95440 Bayreuth, Germany\\
$^{\it 2}$Dept. of Electrical and Computer Engineering, and $^{\it 3}$School of Biomedical Engineering, Colorado State University,
  Fort Collins, CO 80523, USA}

\begin{abstract}
  Diffusion of tracer particles in the cytoplasm of mammalian cells is often anomalous with
  a marked heterogeneity even within individual particle trajectories. Despite considerable
  efforts, the mechanisms behind these observations have remained largely elusive.
  To tackle this problem, we performed extensive single-particle tracking experiments
  on quantum dots in the cytoplasm of living mammalian cells
  at varying conditions.
  Analyses of the trajectories reveal a strong, microtubule-dependent subdiffusion with
  antipersistent increments and a substantial heterogeneity. 
  Furthermore, particles stochastically switch between different mobility states,
  most likely due to transient associations with the cytoskeleton-shaken endoplasmic
  reticulum network.
  Comparison to simulations highlight that all experimental observations can be
  fully described by an intermittent fractional Brownian motion, alternating between
  two states of different mobility.
\end{abstract}

\maketitle

The cytoplasm of mammalian cells is a complex aqueous environment, crowded with large
amounts of macromolecules \cite{fulton1982,Ellis} and a multitude of membrane-enveloped
organelles of largely varying sizes.
Diffusion of supposedly inert tracer particles in the cytoplasm of living cells has
frequently been reported to be anomalous with a sublinear scaling of the mean square
displacement (MSD), $\msd\sim t^\alpha$ ($\alpha<1$) on spatio-temporal scales below a
few micrometers and several seconds \cite{franosch,weiss2014,norregaard2017}. The
emergence of subdiffusive motion appears in many cases to be consistent
with a stochastic process of the fractional Brownian motion (FBM) type
\cite{kepten2011,magdziarz2011,metzler2014}, i.e. a self-similar Gaussian process with
stationary increments whose features are determined by the Hurst coefficient
$H=\alpha/2$ \cite{mandelbrot1968}. FBM dynamics is subdiffusive for $0<H<1/2$ and
trajectories are characterized by antipersistent, i.e. anticorrelated, increments.
A plausible interpretation for such antipersistent memory effects is a viscoelastic
environment \cite{Guigas2007,jed,weber2010,sokolov2012,ernst,krapf2015} with a complex
shear modulus that scales as $G(\omega)\sim\omega^\alpha$, where the elastic and the
viscous parts are responsible for the FBM memory and for energy dissipation, respectively.

Subdiffusion has long been recognized to emerge in solutions crowded with macromolecules,
with an anomaly exponent $\alpha$ that decreases with crowder concentration
\cite{WEK04,banks2005}. However, the value of $\alpha$ is often observed to be
considerably lower in the cytoplasm than in similarly crowded artificially fluids, e.g.
$\alpha\approx0.6$ \cite{Guigas2007,etoc} versus $\alpha\approx0.8$ \cite{jed,ernst}.
Therefore, it is currently understood that subdiffusion in the cytoplasm may not be
caused solely by macromolecular crowding but also relies on additional mechanisms.
As of yet, no general agreement exists for a physical model that can reliably describe
cytoplasmic subdiffusion in detail. Further, subdiffusion is not universal but depends
on tracer size, e.g. for particles in reconstituted entangled actin filament networks,
where $\alpha$ can be continuously tuned between zero and unity as a function of
particle radius and average mesh size \cite{wong2004}. Beyond such caging effects, it
has also been proposed that non-inert crowders may strongly alter the dynamics of
cytoplasmic particles \cite{nagle1992,saxton1996}. Extensive Monte Carlo simulations
have supported this hypothesis \cite{ghosh2015}.
More recently, also experimental support has been obtained via single-particle
tracking (SPT) on surface-modified tracer particles in the cytoplasm of HeLa cells:
The emergence of subdiffusion and the value of $\alpha$ was shown to
depend both on particle size and non-specific interactions to the cytoplasmic interior
\cite{etoc}. Yet, the identity of the cytoplasmic binding partners that enforce the
emergence of subdiffusive motion has remained elusive. Potential candidates include
the cytoskeleton and organelles, e.g. the endoplasmic reticulum (ER) network that
pervades the cytoplasm \cite{JLS}.

Further, local variations in complex media are noticeable in the motion of
particles therein: (Sub)diffusion in cellular fluids has been observed to be 
heterogeneous even within individual trajectories \cite{granick,spako,witzel},
suggesting heterogeneous diffusion processes \cite{cherstvy2013} or spatiotemporal
variations of transport coefficients \cite{slater,chechkin2017,cherstvy2016}.
Despite the elegance of these theoretical models, it remains an open question how a
distribution of apparent diffusivities emerges in the first place. A potential source
might be the ambient active noise in the cytoplasm, i.e. the chemically induced
rattling and shaking of the environment due to the non-equilibrium action of
molecular motors and cytoskeletal filaments. In fact, breaking down cytoskeletal
filaments alters the subdiffusive motion of organelle structures in mammalian
cells \cite{SW2017,SSW2018a} and also compromises the superdiffusive motion
of beads in migrating amoebae \cite{witzel}. Taken together, it is currently neither
clear (i) which mechanism regulates the value of the anomaly exponent $\alpha$ in the
cytoplasm nor (ii) how one should picture the emergence of heterogeneous subdiffusion
due to non-specific interactions in an actively driven environment.

Here, we address these points by extensive SPT experiments on individual quantum dots
loaded into the cytoplasm of living mammalian cells. In particular, we quantify the
particles' motion in the cytoplasm of untreated cells and in cells where the actin
or microtubule cytoskeleton, or the ER has been disrupted. In all cases,
a distinct and heterogeneous subdiffusion of tracers is seen. The subdiffusion effects
become more pronounced when microtubules are broken down. Detailed analyses reveal
that particles switch stochastically between at least two mobility states, irrespective
of the cytoskeleton integrity, but clearly dependent on the presence of an intact ER
network. This evidence suggests non-specific binding of tracers to the ER network,
and hence an indirect coupling to active microtubule-based processes, to be responsible
for the observed heterogeneous subdiffusion in the cytoplasm.
Our experimental data are well described by an intermittent FBM model that switches
stochastically between a higher and lower mobility, supposedly representing free
  motion in the cytosol and co-movement with ER segments.

\begin{figure}[ht] 
  \begin{center}
    \includegraphics[width=8.5cm]{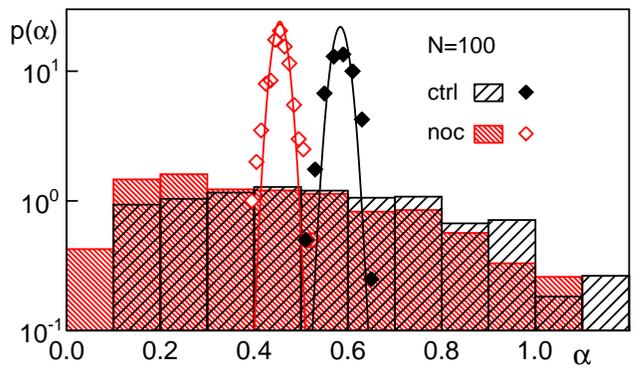}
    \caption{The PDF of anomaly exponents $\alpha$, obtained from individual TA-MSDs
      ($N=100$), shows a broad variation around a mean $\malpha=0.59$ in untreated cells
      (black histogram). Nocodazole-treated cells have a similarly broad PDF (red histogram)
      with a significantly lower mean (cf. Table~\ref{tab01}). Similar results are found
      for longer trajectories (Fig.~S1a in \cite{SM}). Using a bootstrapping
      approach with geometric averaging (diamonds; full lines are Gaussian fits)
      resulted in narrower PDFs with the same mean, $\malpha$.
}
  \label{fig01}
\end{center}
\end{figure}

To explore the heterogeneous subdiffusion in the cytoplasm of mammalian cells, we
performed extensive SPT on quantum dots that had been introduced into the cytoplasm
of cultured HeLa cells by bead loading \cite{mcneil1987glass,SM}. Measurements were
performed with a sampling time of $\Delta t= 100$~ms, and quantum dot trajectories
were first evaluated in terms of their time-averaged MSD (TA-MSD) using $N=100$ or
$N=500$ positions,
\begin{equation}\label{tamsdt}
   \msdt= \frac{1}{N-k}\sum_{i=1}^{N-k}
  \left[{\bf r}((i+k)\Delta t)-{\bf r}(i\Delta t)\right]^2 \,\,.
\end{equation}
Following previous reports \cite{etoc,manzo2015,weron2017}, individual TA-MSDs were
fitted with a simple power law $\msdt=K_\alpha\tau^\alpha$ in the range
$\Delta t\le\tau\le 10\Delta t$ to extract the anomaly exponent $\alpha$ and the
generalized diffusion coefficient $K_\alpha$. The resulting probability density
function (PDF) of anomaly exponents, $p(\alpha)$, showed considerable
trajectory-to-trajectory fluctuations around a mean $\malpha\approx0.57$ (\fig{fig01}
and Fig.~S1a in \cite{SM}) that slightly depends on the trajectory length $N$
(Table~\ref{tab01}).
Control experiments in highly viscous artificial solutions yielded
  $\malpha\approx1$ (Sect.~C and Fig.~S2f,g in \cite{SM}).
%

To probe a potential perturbation of the power-law scaling due to static and dynamic
localization errors \cite{moerner}, and to validate the significance of the mean
exponent $\malpha$, we exploited a bootstrapping approach \cite{SM}: From the whole set
of calculated TA-MSDs we drew randomly a non-exhaustive ensemble of $100$ curves,
averaged these geometrically, and used again a simple power-law fit to extract the
scaling exponent $\alpha$ of the resulting ensemble-averaged TA-MSD. Repeating this
approach $M=200$ times, we noted that none of the ensemble-averaged TA-MSDs showed a
significant offset in the limit $\tau\to0$
(Sect.~E and Fig.~S2a-d in \cite{SM}). Hence, positive and negative contributions
from static and dynamic localization errors
appear to cancel each other in our data and therefore fitting with a simple power law gives
meaningful results for $\alpha$.

The PDF of $\alpha$ values obtained with the bootstrapping approach (\fig{fig01}) 
was very narrow  with a mean $\malpha$ that matched the respective value found
before via individual TA-MSDs (Table~\ref{tab01}). Geometric averaging of TA-MSDs boils
down to an arithmetic averaging of individual $\alpha$ values (but not of $K_\alpha$).
Thus, the narrow width of $p(\alpha)$ after bootstrapping is determined by
$\sigma/\sqrt{M}$, where $\sigma$ is the standard deviation of $\alpha$ derived from
individual TA-MSDs.
Analyzing TA-MSDs with a recently introduced and validated resampling algorithm
\cite{weiss2019} confirmed the values for $\malpha$ \cite{SM}.
An arithmetic instead of a geometric averaging of TA-MSDs lead to an overestimation of the
mean scaling exponent (Table~\ref{tab01} and Fig.~S1b in \cite{SM}).  
\begin{table}[t]
\begin{centering}
\begin{tabular}{l|c|c|c|c}
                   & untreat. & noc & cyto~D & lat~A \\[1mm] 
  \hline \\[-2mm]
  TA-MSDs & $0.59$ ($0.55$) & $0.46$ ($0.36$) & $0.58$ ($0.54$) & $0.62$ ($0.58$) \\[1mm]
  b.tr. geom.  & $0.58$ ($0.55$) & $0.46$ ($0.36$) & $0.58$ ($0.54$) & $0.61$ ($0.57$) \\[1mm]
  b.tr. arith. & $0.79$ ($0.60$) & $0.66$ ($0.43$) & $0.82$ ($0.73$) & $0.86$ ($0.76$) \\[1mm]
  \hline
\end{tabular}
\par\end{centering}
\caption{Mean anomaly exponents $\malpha$ for trajectories of length $N=100$
  ($N=500$) in untreated cells and after application of nocodazole, cytochalasin~D, or
  latrunculin~A. Standard errors were in all cases smaller than $0.02$.} \label{tab01}
\end{table}

Being interested in how cytoplasmic diffusion is affected by the cytoskeleton, we
applied either nocodazole to break down microtubules, or cytochalasin~D or latrunculin~A
to disrupt actin filaments. Disrupting microtubules changed the diffusion anomaly
substantially (\fig{fig01} and Table~\ref{tab01}) whereas disrupting actin networks
had no significant effect (Table~\ref{tab01}). Transport coefficients $K_\alpha$ showed
a higher sensitivity to microtubule disruption and also a stronger dependence on
trajectory length (Fig.~S1c in \cite{SM}). Similar to previous observations on the
dynamics of the ER \cite{SSW2018a}, the effect of nocodazole on $K_\alpha$ was not
particularly strong for short trajectories. For longer trajectories, however, a marked
shift to smaller transport coefficients was visible upon microtubule disruption. This
puts up a caveat that longer trajectories may represent a distinct subset of the
acquired data, e.g. a lower mobility facilitating the tracking, but it also indicates
that microtubule-associated processes significantly contribute to the diffusion anomaly
in untreated cells beyond a change in the scaling of MSDs.

\begin{figure}[ht] 
  \begin{center}
    \includegraphics[width=8.5cm]{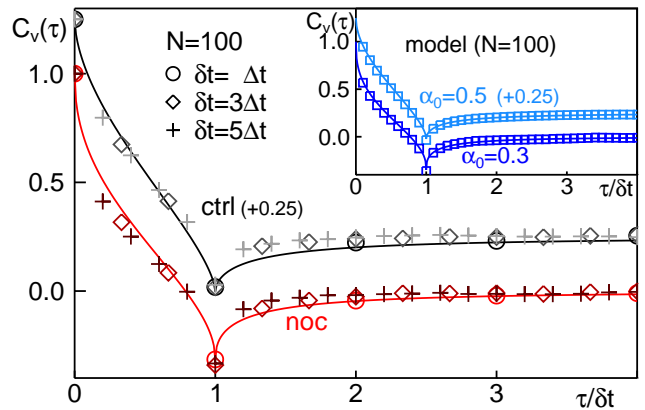}
    \caption{Rescaled normalized VACFs [\Eq{vacf}] of all experimental trajectories
      with $N=100$ at different $\delta t$ agree with the analytical prediction for
      FBM (full lines, \Eq{vacfFBM}), without treatment (grey symbols) and after
      nocodazole-treatment (red symbols). For better visibility, untreated cell data have
      been shifted upwards. No significant differences are seen for longer trajectories
      (Fig.~S3a in \cite{SM}).
      An estimate of $\malpha$ can be directly obtained from the VACF minimum,
        $C_v(\xi=1)=2^{\alpha-1}-1$. VACF minima for untreated and nocodazole-treated cells
        yield $\alpha=0.58\pm0.01$ and $\alpha=0.38\pm0.02$, respectively, in favorable
        agreement with our MSD results.
      Inset: VACFs of simulated intermittent FBM trajectories
      ($N=100$, anomaly parameter $\alpha_0$) also agree with \Eq{vacfFBM} (full lines).}
  \label{fig02}
\end{center}
\end{figure}

Going beyond the MSD, we analyzed the ensemble average of the velocity autocorrelation
function (VACF), 
\begin{equation}\label{vacf}
C_v(\tau)=\langle {\bf v}(t){\bf v}(t+\tau)\rangle_{t,E} 
\end{equation}
that is highly sensitive to the nature of unconfined anomalous diffusion processes
\cite{burov2011,weber2012}. Here,
${\bf v}(t)=[{\bf r}(t+\delta t)-{\bf r}(t)]/\delta t$ is the velocity at time $t$,
given via the increments in a period $\delta t$. Varying $\delta t=k\Delta t$ in
multiples of the sampling time $\Delta t$, the VACFs showed in all cases a pronounced
negative peak for $\tau=\delta t$ as expected for antipersistent random walks. By
rescaling the times as $\xi=\tau/\delta t$, all VACF traces collapse to a single master
curve that agrees with the analytical predictions for FBM (\fig{fig02} and Fig.~S3a in
\cite{SM}), namely
\begin{equation}\label{vacfFBM}
  C_v(\xi)=\left\{(\xi+1)^\alpha+|\xi-1|^\alpha-2\xi^\alpha\right\}/2\,\,,
\end{equation}	
with $\alpha$ being set to the value $\malpha$ found with the bootstrapping protocol
(Table~\ref{tab01}). We emphasize the exceptional agreement of the experimental data
with \Eq{vacfFBM} without any fitting parameters since other antipersistent random walk
data, e.g. from membrane proteins, can deviate significantly from the FBM prediction
(see Fig.~S3b \cite{SM} for an example).
%

\begin{figure}[ht] 
  \begin{center}
    \includegraphics[width=8.5cm]{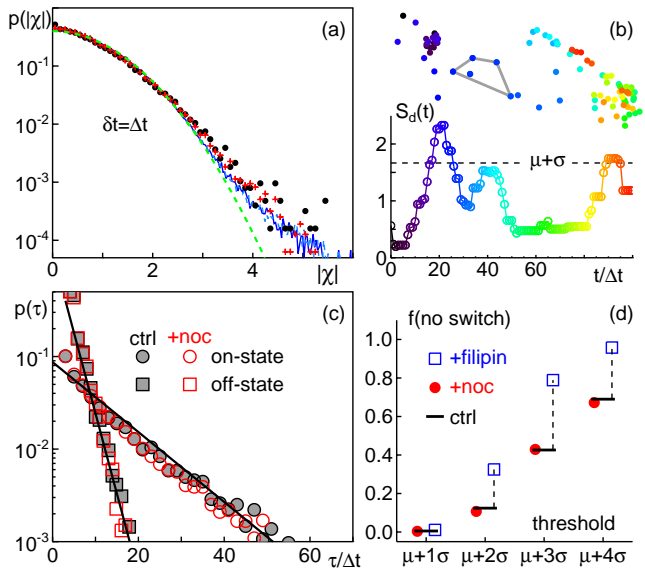}
    \caption{(a) PDFs of normalized increments (time lag $\delta t=\Delta t$, shown
        here as moduli, $|\chi|$) follow the anticipated Gaussian (green dashed line) for
        small $|\chi|$ but show significant deviations for $|\chi|>3.5$, indicating a heterogeneous
        process (black circles and red crosses: untreated and nocodazole-treated cells). These
        data are
        in excellent agreement with simulations of an intermittent FBM model ($\alpha_0=0.5$
        and $\alpha_0=0.3$: coinciding light and dark blue lines). (b) Representative trajectory
        (color-coded successive positions) with a
        local convex hull (LCH) at $t=28\Delta t$ highlighted in grey. The corresponding
        time series of largest LCH diameters, $S_d(t)$, shows considerable fluctuations.
        Values $S_d(t)\ge\mu+\sigma$ (dashed horizontal line) are rated to be in the
        more mobile 'off'-state. (c) Residence times in the low- and high-mobility state,
        extracted from individual trajectories (threshold $\theta=\mu+\sigma$) feature
        exponential PDFs (full black lines) with a substantially longer mean residence time
        in the 'on'-state. No
        substantial differences are seen for nocodazole-treatment or when choosing a threshold
        $\theta=\mu+2\sigma$ (Fig.~S6c in \cite{SM}).
        (d) The fraction of trajectories without any switching rises when successively increasing 
        the threshold value to $\theta=\mu+4\sigma$. No significant differences
        are seen between untreated (black horizontal stripes) and nocodazole-treated cells
        (filled red circles). In contrast, trajectories from filipin-treated cells (open blue
        squares) feature a much stronger increase (highlighted by dashed lines), indicating
        that ER structures are required for the mobility switching.
}
  \label{fig03}
\end{center}
\end{figure}

Next we inspected the PDF of the normalized increments $\chi$ within a time lag
$\delta t$ \cite{spako}, i.e. time series $\Delta x_i=x_{i+k}-x_i$ and $\Delta y_i=y_{i+k}-y_i$
were calculated and normalized by their individual root-mean-square step length.
Since no systematic differences were observed between $x$- and $y$-directions, all
normalized increments were combined into a single set of $\chi$. For a homogeneous
FBM, a Gaussian PDF $p(\chi)$ is expected for all $\delta t$. Yet, for small $\delta t$
our data showed significant deviations from a Gaussian in the tails of the distribution
(\fig{fig03}a and Fig.~S4a in \cite{SM}). This suggests that individual trajectories are heterogeneous,
i.e. the particle mobility changes within the trajectory. For $\delta t=10\Delta t$, this
heterogeneity subsides, collapsing the increment statistics to the anticipated Gaussian
(Fig.~S4b in \cite{SM}).
%

To directly probe switching between different mobilities, we analyzed the local
convex hull (LCH) of individual trajectries \cite{lch,SM}: After normalizing the trajectories
by their root-mean-square step length, we determined for each trajectory the largest diameter
$S_d(t)$ of the LCH for positions visited in the period $[t-2\Delta, t+2\Delta]$ (see \fig{fig03}b
for illustration). Using the mean $\mu$ and standard deviation $\sigma$ of all $S_d$ values
for a given cell condition, we defined a threshold $\mu+\sigma$ and rated particles to
be in a more mobile state for $S_d(t)\ge \mu+\sigma$ (see also Fig.~S6 in \cite{SM}). 
As a result, we observed a frequent switching between a lower- and a higher-mobility state
(named 'on' and 'off', respectively) with markedly larger mean residence times $\tau$ in
the low-mobility state, irrespective of any treatment (see PDFs $p(\tau)$ in \fig{fig03}c).
Employing a threshold $\mu+\sigma$, all trajectories exhibit switching behavior. However,
upon increasing the threshold, a growing fraction of trajectories does not display any switching
(\fig{fig03}d) while no substantial difference is seen in the mean residence times (see
Fig.~S6c in \cite{SM}). Hence, the LCH analysis
confirms the existence of at least two mobility states for untreated and nocodazole-treated
cells. Additional support for a switching behavior is given by the autocorrelation function
of squared increments, $G(\tau)$, which shows a long-lasting decay (Sect.~G and Fig.~S5 in
\cite{SM}).
\begin{figure}[ht] 
  \begin{center}
    \includegraphics[width=8.5cm]{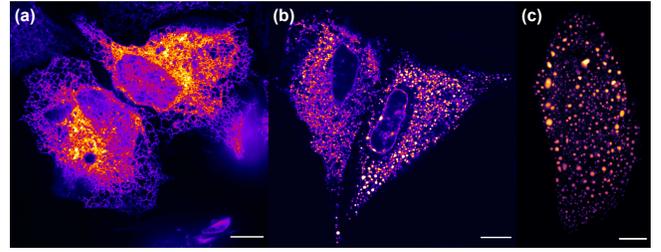}
    \caption{Representative fluorescence images of the ER in (a) untreated, (b) filipin-treated,
       and (c) osmotically shocked cells. In line with previous reports \cite{axelsson2004,JLSer},
       the ER network of untreated cells is completely fragmented after filipin treatment or
       osmotic shock. Scale bars: $10~\mu$m.}
  \label{fig04}
\end{center}
\end{figure}

Based on these results and previous observations on the cytoskleton-dependent anomalous
dynamics of ER junctions \cite{SSW2018a}, we hypothesized that particle interactions with the
ubiquitous ER network are key for the observed switching of mobilities. We therefore repeated
tracking experiments in cells where the ER network had been fragmented either using
the drug filipin \cite{axelsson2004,SM} or by an osmotic shock \cite{JLSer,SM} (see \fig{fig04}).
We observed that a lack of ER tubules did not grossly alter the
scaling exponent $\malpha\sim0.6$. Yet, the intermittent nature of the particle motion was
markedly reduced as evidenced by the LCH analysis (\fig{fig03}d), i.e. trajectories with a
switching of mobilities become more rapidly diminished when altering the threshold for $S_d(t)$.
This indicates that association to and dissociation from ER tubules is involved in
creating the intermittent nature of the particles' diffusion.

Having observed a heterogeneous, intermittent, ER- and cytoskeleton-dependent
subdiffusion of quantum dots in the cytoplasm, we used Occam's razor to formulate the
simplest model that can capture our experimental data (see \cite{SM} for a discussion of
more elaborate models). Taking all experimental constraints into account, we arrived at an
intermittent FBM model: We modeled the dynamics of individual particles as FBM with fixed
anomaly $\alpha_0$ and a transport coefficient that randomly switches within each trajectory
\cite{SM}. Particles were assumed to exist in 'on' and 'off' states with transport
coefficients $K^{\text{on}}_\alpha <K^{\text{off}}_\alpha$, representing ER-tubule
associated and free motion. Dichotomous switching between these states was modeled as a
Markov process with transition rates $k_{\text{on}}$ and $k_{\text{off}}$. In our simulations we
kept these rates and the ratio $s=K^{\text{on}}_\alpha /K^{\text{off}}_\alpha$ fixed, and chose
$\alpha_0=0.5$ ($\alpha_0=0.3$) for untreated (nocodazole-treated) cells, in accordance with
the previously reported anomaly values for ER junctions \cite{SSW2018a}. Despite the simplicity
of this model, we observed a surprisingly good overlap with our experimental data when choosing
$s=3.5$, $k_{\text{on}}=0.27$~s$^{-1}$, and $k_{\text{off}}=0.01$~s$^{-1}$: First, the
mean anomaly of simulated realizations, extracted from TA-MSDs, was $\malpha=0.55$ and
$\malpha=0.37$, respectively, in agreement with experimental observations
(Table~\ref{tab01}). The slightly larger value as compared to the imposed value $\alpha_0$ is
a consequence of the dichotomous switching that perturbs the pure FBM behavior. Second, when
using the respective value $\malpha$, the VACF showed the same agreement with \Eq{vacfFBM} as
the experimental data (insets of \fig{fig02} and Fig.~S3a in \cite{SM}). Third, the non-Gaussian
shape of the increment $\chi$ statistics for $\delta t=\Delta t$ and a more Gaussian shape
for $\delta t=10\Delta t$ are almost perfectly matched (\fig{fig03}a and Fig.~S4 in \cite{SM}).
Fourth, the shape of $G(\tau)$ overlapped very well with the experimental data (Fig.~S5 in
\cite{SM}). Moreover, the PDFs of residence times in the 'on' and 'off' states were in
favorable agreement with our experimental results (Sect.~F in \cite{SM}). We therefore conclude
that our minimal model is sufficient for reproducing the features of our experimental data.

In summary, we have observed a heterogeneous and intermittent subdiffusion of quantum dots
in the cytoplasm of living cells that was altered upon disrupting microtubules or fragmenting
the ER network. Our experimental data are well described by a simple intermittent FBM model in
which we have set the anomaly exponents to those observed for the motion of ER junctions in
untreated and nocodazole-treated cells. Combining all insights, we arrive at the conclusion
that transient association with ER membranes hampers free diffusion of the particles, hence
enforcing a particularly low anomaly exponent $\alpha$. If the ER network is intact, association
with ER tubules leads to an intermittent diffusion process of particles and couples their
motion indirectly to active microtubule-based processes. The persisting, strongly subdiffusive
type of motion after fragmenting the ER network hints at additional structures with which
particles might interact, e.g. networks of intermediate filaments \cite{intermed_fil} or
(ER-derived) membrane vesicles that mimic a microemulsion \cite{hellweg}. Thus, subdiffusion
in the cytoplasm is indeed a considerably more complex phenomenon than anomalous diffusion in
artificial fluids crowded with passive macromolecules.

\begin{acknowledgments}
We thank Mike Tamkun for discussions on experimental design, O'Neil Wiggan for providing
HeLa cells, and Ashok Prasad and Wenlong Xu for providing the bead loader assembly and
helping with the  bead loading protocol. This work was financially supported by the
German Academic Exchange Service (PPP USA grant No. 57315749). AS and MW also
gratefully acknowledge financial support by the VolkswagenStiftung (Az. 92738).
\end{acknowledgments}


\end{document}